\documentclass{aa}

\usepackage{graphicx}

\usepackage{txfonts}

\usepackage{amsmath}	
\usepackage{multicol}        
\usepackage{bm}		
\usepackage{pdflscape}	
\usepackage{amsfonts}
\usepackage{amssymb}
\usepackage{physics}
\usepackage{epsfig} 
\usepackage{xcolor}
\usepackage{url}
\usepackage{soul}
\usepackage{hyperref}

\newcommand{\f}{{\textit{f}}}
\newcommand{\ms}{~m~s$^{-1}$}
\newcommand{\kms}{~km~s$^{-1}$}

\newcommand{\flows}{\mathrm{f}} 
\newcommand{\geom}{\mathrm{g}} 

\begin{document} 
   \title{Coriolis force acting on near-surface horizontal flows during simulations of flux emergence produces a tilt angle consistent with Joy’s law on the Sun}

    \author{
    W.~Roland-Batty$^{1}$,
    H.~Schunker$^{1}$\thanks{E-mail: hannah.schunker@newcastle.edu.au},
    R.H.~Cameron$^{2}$,
    D.~Przybylski$^{2}$,
    L.~Gizon$^{2,3}$
    and D.I.~Pontin$^{1}$
    \\
    $^{1}$The University of Newcastle, University Dr, Callaghan NSW 2308, Australia\\ 
    $^{2}$Max-Planck-Institut f\"{u}r Sonnensystemforschung, 37077  G\"{o}ttingen, Germany\\
    $^{3}$Institut f\"{u}r Astrophysik, Georg-August-Universit\"{a}t G\"{o}ttingen, 37077 G\"{o}ttingen, Germany
    }

   \date{Received September 15, 1896; accepted March 16, 1897}

 
  \abstract
   {Joy’s law describes the tilt of bipolar active regions on the Sun away from an east-west orientation, where the flux of the polarity concentrated at the prograde side tends to be closer to the equator than the polarity on the retrograde side. Joy's law is attributed to the Coriolis force because of the observed increase in tilt angle at higher latitudes. This tilt plays a crucial role in some solar dynamo models.}
   {Our goal is to model the effects of the Coriolis force on a flux tube as it rises through the near-surface convection zone.  }
   {We use a three-dimensional Cartesian magnetohydrodynamic simulation of an untwisted flux tube ascending from a depth of 11~Mm. We model the Coriolis effect using the \f-plane approximation, that only considers and acts on horizontal flows. On the Sun, Joy's law is weak and is only evident as an average over many active regions. To achieve a measurable effect in a single simulation, we consider a rotation rate $110$ times faster than the Sun. 
    }
   {The simulation shows that the flux tube emerges at the surface with a tilt angle consistent with Joy’s law when scaled to the Sun's slower rotation, and the tilt angle does not substantially change after emergence.}
   {This shows that the Coriolis force acting on flows horizontal to the surface within the near-surface convection zone is consistent with Joy’s law.}
    \titlerunning{Near-surface simulations of Joy's Law}
    \authorrunning{W. Roland-Batty et al.}
   \maketitle
%

\section{Introduction}
The orientation of magnetic active regions on the solar surface obeys Hale's law and Joy's law \citep{Haleetal1919}.
Simple active regions consist of a bipole pair aligned in a predominantly east-west  orientation.
Hale's law states that the sign of the leading (relative to the Sun's rotation) polarity in each hemisphere are opposite, and the sign of the polarities switches from one solar cycle to the next.  
Joy's law describes the statistical tendency of the leading polarity (in the direction of the Sun's rotation) of an active region to be closer to the equator than the following polarity. Studies have shown that the tilt of the polarities increases with (absolute) solar latitude \citep{WangSheeley1989}. These two laws play a crucial role in the Babcock-Leighton dynamo model \citep{Babcock1961,Cameronetal2015,KarakMiesch2017}.

Joy's law has conventionally been quantified using continuum intensity observations of sunspots \citep[e.g.][]{McClintockNorton2016}, which limits the description to well-developed and stable active regions, with established sunspots. To pin-point the origin of active region tilt angles, however, observing the magnetic field is required to capture the earliest stages of flux emergence. 

Monitoring missions, such as the Michelson Doppler Imager (MDI) onboard the Solar and Heliospheric Observatory \citep[SoHO][]{Scherrer1995} and the Helioseismic and Magnetic Imager \citep[HMI][]{Schou+2012}  onboard the Solar Dynamics Observatory \citep[SDO][]{SDO2012}, make routine observations of the Sun's surface magnetic field at a high cadence and with a high duty-cycle that are able to capture the early stages of active region emergence. From over 700 active regions \citet{KosovichevStenflo2008} found that, statistically, active regions emerge roughly east-west aligned with zero tilt angle. By analysing the motion of the individual polarities in SDO/HMI line-of-sight magnetic field observations, \citet{Schunker+2020} found that the tilt angle develops during the emergence process \citep[1-2~days,][]{Weberetal2023} and that the latitudinal dependence of the tilt angle is due to the north-south separation of the polarities. 
The latitudinal dependence suggests that the Coriolis force is responsible for this north-south motion.  That raises the question of which flows are influenced by the Coriolis force to generate the tilt angle.

One possibility involves the turbulent convective velocities \citep{Parker1955,ChoudhuriDSilva1990,Brandenburg2005}. \cite{Schmidt1968} proposed that active region bipoles emerge with upwelling supergranulation cells, where the surface flows within the cell push the polarities outwards, further separating them.
An alternative explanation arises from the buoyant ascent of magnetic flux tubes through the Sun's convection zone \citep{WangSheeley1991,DSilvaChoudhuri1993,Fisher+1995,Weberetal2011}. These studies usually employ the thin-flux-tube approximation that is invalid in the top $\sim 20$~Mm beneath the surface \citep[see][and citations within]{Fan2009}.

The north-south component of the Coriolis force is associated with the creation of radial vorticity. This radial vorticity can introduce a north-south separation of the two magnetic polarities during flux emergence, giving rise to Joy's law. In this paper, we investigate the effect of the radial vorticity generated by the Coriolis force on the orientation of an emerging flux tube. For this purpose we consider the effect of the Coriolis force acting on horizontal flows (equivalently, the radial vorticity component).

 We simulate a flux tube rising through the near-surface convective region of the Sun and then through the surface. In Section~\ref{sec:muram} we introduce our numerical approach, including a description of the implementation of the \f-plane approximation of the Coriolis force, and the initial condition for the flux emergence simulations. In Section~\ref{sec:tiltangle}, we describe the evolution of the tilt of the polarities in each simulation in relation to the flows. We discuss the implications of our results and conclusions in Section~\ref{sect:conc}.

\section{Numerical Approach}\label{sec:muram}

To simulate the interaction of a flux tube rising through the near-surface convection of the Sun, we use the Max-Planck-Institute for Aeronomy/ University of Chicago Radiation Magneto-hydrodynamics (MURaM) code  (\citet{Vogler+2005}; and further developed by \citet{Rempel2014,Rempel2017}). This code solves the  magnetohydrodynamics (MHD) equations in a three-dimensional Cartesian box that includes a section of the surface of the Sun. The local plasma velocity, $\mathbf{v}$, evolves with the density, $\rho$, pressure, $p$, and magnetic flux density, $\mathbf{B}$, through the MHD equations:

\begin{equation}
    \pdv{\rho}{t} = -\nabla\cdot(\rho\mathbf{v})
\end{equation}
\begin{equation}
\begin{split}
      \pdv{\rho\mathbf{v}}{t} &=& -\nabla\cdot(\rho\mathbf{vv})-\nabla p +  \rho \mathbf{g} +\mathbf{F}_\text{L} + \mathbf{F}_\text{SR} + \rho\mathbf{a}_\mathrm{cor}
     \label{eqn:motion1}
\end{split}
\end{equation}
\begin{equation}
\begin{split}
     \pdv{E_{HD}}{t} &= &-\nabla\cdot\left[\mathbf{v}(E_{HD}+p)
     \right]+\rho\mathbf{v}\cdot (\mathbf{g} +\mathbf{a}_\mathrm{cor})\\
    &\quad& + \mathbf{v}\cdot(\mathbf{F}_\text{L}+\mathbf{F}_\text{SR})+Q_{\text{rad}}  
    \label{eqn:e1}
\end{split}
\end{equation}

\begin{equation}
    \pdv{B}{t}=\nabla\times(\mathbf{v}\times\mathbf{B}) 
\end{equation}
with time, $t$, and temperature, $T$; 
$E_{HD}= E_{\mathrm{int}} + \frac{\rho \mathbf{v}^2}{2}$, the sum of the internal and kinetic energy; and $\mathbf{F}_\text{SR}$ limits the Alfv\'{e}n speed  \citep[for more detailed definitions see][]{Rempel2014,Rempel2017}. 
We added the Coriolis acceleration, $\mathbf{a}_\mathrm{cor}$,  to the equations solved by MURaM which we describe in Section.~\ref{sec:fplane}. We note that since $\mathbf{a}_\mathrm{cor}$ and $\mathbf{v}$ are perpendicular, this term has no effect on the energy.

The Lorentz force is given by
$$
\mathbf{F}_\text{L} = \frac{f_{A}}{4\pi}\nabla\cdot\left(\mathbf{BB}-\frac12\mathbf{I} B^2\right) + \frac{(1-f_A)}{4\pi}(\nabla\times\mathbf{B})\times\mathbf{B},
$$
$$
\mathrm{where}\quad f_A = \frac{1}{\sqrt{1+\left(\frac{v_A^4}{c^4}\right)}}, 
$$
the Alfv\'{e}n velocity is $v_A$, and the identity tensor is $\mathbf{I}$. The parameter $c$  limits the speed of Alfv\'{e}n waves, implemented specifically for extending the atmosphere of MURaM into the chromosphere and corona. The case when $c$ is set to the the speed of light corresponds to relativistic Alfv\'{e}n waves. We limit the Alfv\'{e}n wave speed to 100~km~s$^{-1}$ which should be inconsequential in our simulations of the convection zone and photosphere.

We use a constant value for the gravitational acceleration $\mathbf{g}=-2.74 \times 10^4$~cm$^2$s$^{-1}$ in the vertical direction.

In the energy equation (Eqn.~\ref{eqn:e1}), $Q_\mathrm{rad}$ is the radiative heating (and cooling) term. The radiation transport equations are solved using the short characteristics method \citep{KunaszAuer1988,Vogler+2005}. We treat the radiation as monochromatic (grey) with opacities from the MPS ATLAS package \citep{witzke+2021}. We define the solar surface in the simulation to be the $\tau_\mathrm{500}=1$ layer.

An equation of state is required to close the system. 
We assume the plasma is in local thermodynamic equilibrium. 
The equation of state is determined numerically by relations specifying $T (\rho, E_\mathrm{int})$ and $p(\rho, E_\mathrm{int})$ where $E_\mathrm{int}$ is the internal energy density, which are read from tables generated using the FreeEOS package \citep{Irwin2012}.

MURaM has a free parameter, $h$, that controls the numerical diffusion  \citep{Rempel2014}. Following  \citet{Rempel2017} we use a value of $h=2$ for the lower boundary and regions with density  $\rho>10^{-11}$~g/cm$^3$ (mostly below the surface), $h=1$ for regions with density below $\rho \leq 10^{-11}$~g/cm$^3$ (above the surface), and $h=0$ for the upper boundary. A discussion of the implications of the $h$ parameter and diffusion scheme is given in \cite{Rempel2017}.

The top boundary condition of the domain is semi-transparent, permitting outflows but not inflows, and the magnetic field is forced to be purely vertical. 
We use the `Open boundary/symmetric field' condition \citep[\textit{OSb.} in][]{Rempel2014} at the bottom boundary which is open to flows and is symmetric for the magnetic field. We set a constant entropy in the inflows at the bottom boundary and  extrapolate the pressure across the boundary cells.
The horizontal boundaries are periodic.

With MURaM, we simulated the near-surface of the Sun from 15.128~Mm below the surface to 1~Mm above (a total of 504 grid points with a resolution of 0.032~Mm in the $z$-direction), and extending 96.768~Mm (over 1008 grid points with a resolution of 0.096~Mm) in each horizontal direction $(x, y)$. 

To initiate the simulations, we prescribed the density, temperature, sound-speed and pressure of a standard solar model \citep[based on Model S]{JCDetal1996} in the box that varies only in height, and initiated the convection with small amplitude random velocity perturbations uniformly distributed between $\pm 100$~cm~s$^{-1}$. Importantly, we initiated the convection without rotation, i.e. $\mathbf{a}_\mathrm{cor}=0$. Starting from our initial condition the simulation required about 8 solar hours until convection reached a steady state. After reaching this steady state, we selected a single snapshot as the initial condition for all following simulations.

\subsection{Implementing the Coriolis force in the \f-plane approximation}
\label{sec:fplane} 
Our goal is to understand the effects of the radial vorticity introduced by the Coriolis force.
The Coriolis force is a pseudo-force felt by objects in a rotating frame of reference. The Coriolis acceleration is given by 2$\mathbf{\Omega}\times$\textbf{v} where $\mathbf{\Omega}$ is the rotation vector and \textbf{v}$=(v_\phi,v_\theta, v_r)$ is the velocity within the reference frame in spherical coordinates  where $\theta$ is latitude, $\phi$ is longitude, and $r$ is the radial direction. 
The Coriolis acceleration is the cross product of the flow velocity, and the rotation rate,
\begin{equation}
    2\mathbf{\Omega}\times\mathbf{v}=2(v_r\Omega\cos\theta-v_\theta\Omega\sin\theta, v_\phi\Omega\sin\theta, -v_\phi\Omega\cos\theta),
\end{equation}
where the first term is in the longitudinal direction, $\boldsymbol{\hat{\phi}}$, the second term is in the latitudinal direction, $\boldsymbol{\hat{\theta}}$, and the third is in the radial direction, $\mathbf{\hat{r}}$.
In our case, we are simulating a localised box covering a small section of the Sun's surface that we consider to be located at a single latitude.  We approximate the plasma velocity in spherical coordinates to Cartesian coordinates in our box by $(v_\phi, v_\theta, v_r) \rightarrow (v_x, v_y, v_z)$. 

Because we are interested in the radial vorticity introduced by the Coriolis force, we consider the terms generating radial vorticity from horizontal flows via Coriolis acceleration in the \f-plane approximation
\begin{equation}
    \mathbf{a}_\mathrm{cor} = 2 \mathbf{\Omega} \times \mathbf{v} = (-f v_y,f v_x, 0) \mathrm{  \quad  where \quad  } f=2 \Omega \sin \theta.
     \label{eq:fplane}
\end{equation}
We plan to include the effects of the other terms in the \f-plane approximation in a future paper.

In our simulations we chose a  mid-latitude of solar activity in the Northern hemisphere, $\theta=15^\circ$, and a rapid rotation rate of $\Omega = 47.75~\mu$~Hz $\approx 110 \, \Omega_\odot$ (where the Sun has a surface rotation rate of about 2~kms$^{-1}$, or 430~nHz). 

We selected a fast rotation rate to ensure we would have a measurable effect from the Coriolis force on the typical convective velocities in a practical time-frame on the order of about 10~hours of solar time. We ignore any perturbations to the asphericity or local gravity, since we are only interested in the effect of the Coriolis force on the horizontal flows, as in the case for the Sun.

We added the Coriolis term to the magnetohydrodynamic equation of motion (Eqn.~\ref{eqn:motion1}) and energy equation in the MURaM code (Eqn.~\ref{eqn:e1}). The local plasma velocity then evolves with the MHD equations, with the added effect from the rotation in the \f-plane approximation with the $\mathbf{a}_\mathrm{cor}$ term.

To quantify the effect of the Coriolis force on the supergranulation scale convective flows, we simulated the behaviour of the convection with, and without, the Coriolis acceleration for $\approx 24$~hours starting from identical initial conditions.  We refer to the simulations  with $\Omega = 110 \, \Omega_\odot$ as the `rotational simulation', and $\Omega = 0 \, \Omega_\odot$ as the `non-rotational simulation'.

\begin{figure}
    \centering
    \includegraphics[width=0.5\textwidth]{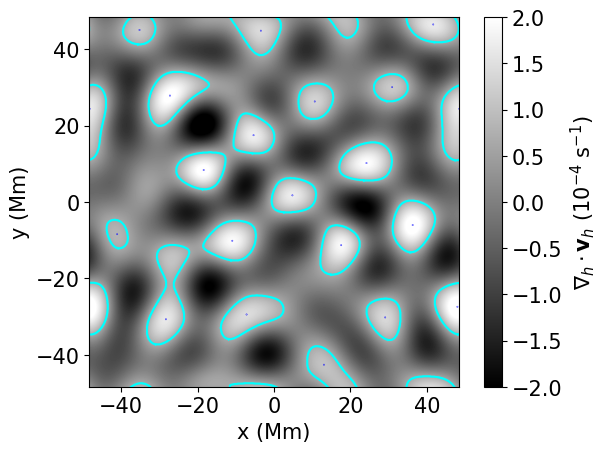}
    \vspace{-0.5cm}
    \caption{Example of a surface horizontal flow divergence map averaged over 6~hours  and filtered between 15~Mm and 25~Mm from the non-rotational hydrodynamic simulation. The cyan contours indicate the identified supergranules and the blue points indicate the location of the peak divergence within the boundary.}
    \label{fig:divmap}
\end{figure}

We validated the implementation of the \f-plane approximation by comparing the divergence and vorticity of the flows in the hydrodynamic simulation to an analytic estimate (see Appendix~\ref{app:valfplane}). A visual representation of the effects of the Coriolis force on the flows can be seen in surface supergranulation flows. We averaged the $v_x$ and $v_y$ surface velocity maps in 6-hour-long segments 
and filtered out spatial scales less than 15~Mm and greater than 25~Mm (the approximate size of supergranules which could form in the limited size of the simulation box). From each of these four segments we computed maps of the horizontal flow divergence, $\nabla_h \cdot \mathbf{v}_h$ (see Fig.~\ref{fig:divmap}).

To identify supergranules, in each temporally averaged map of flow divergence we set a threshold of 0.8-standard deviations from the mean of all four flow divergence maps, following the definition used by \cite{Hirzberger+2008} and \cite{Langfellneretal2015_Vorticity}. We contoured the boundaries of the features at the threshold value, and defined the centre of each feature as the pixel with peak horizontal divergence within the boundary. If the peak lies less than 1~Mm distance from any part of the contour  we exclude it from our selection of supergranules (see the example in Fig.~\ref{fig:divmap}).

\begin{figure}
    \centering
    \includegraphics[width=0.99\linewidth]{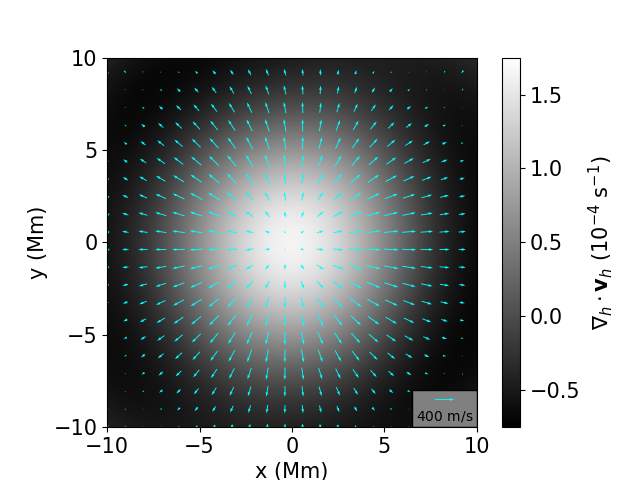}
    \includegraphics[width=0.99\linewidth]{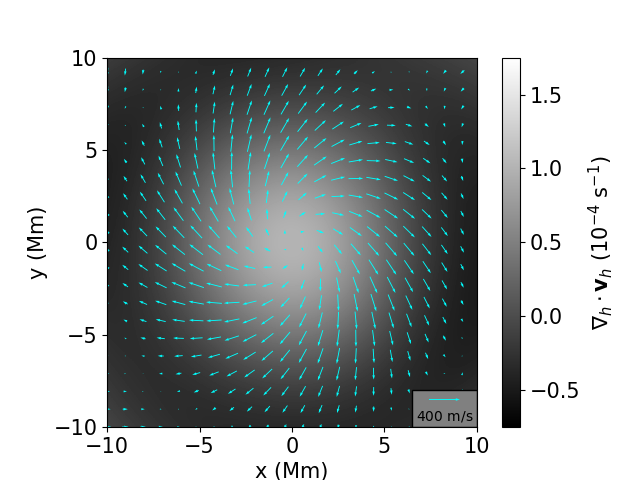}
    \includegraphics[width=0.99\linewidth]{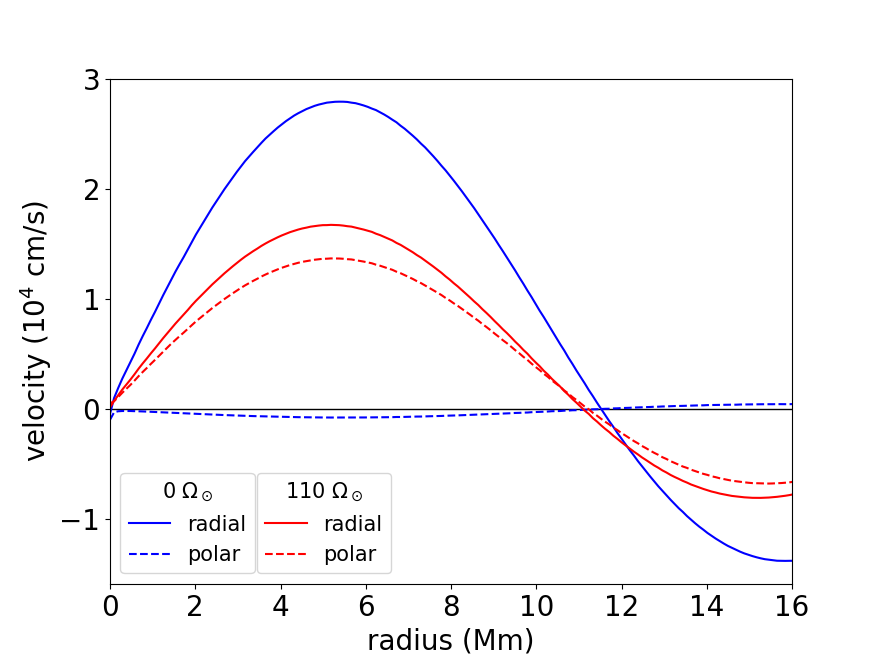}
    \caption{Top panel: Horizontal flow divergence of the average supergranule in the non-rotational hydrodynamic simulation. The cyan arrows represent the horizontal flows at the surface. 
    Middle panel: same as for the top panel but for the rotational hydrodynamic simulation. 
    Bottom panel: azimuthal averages of the radial velocity ($(x v_x + y v_y)/\sqrt{x^2+y^2}$) and polar velocity ($(y v_x-x v_y)/\sqrt{x^2+y^2}$ )  of the flows for the average supergranule. All quantities are measured relative to the centre of the supergranule. The polar velocity is defined as positive in the clockwise direction.}
    \label{fig:aveSGdiv}
\end{figure}

We averaged the horizontal flow divergence maps centred at each supergranule peak to create the average supergranule (see Fig.~\ref{fig:aveSGdiv}, top). We repeated this for the horizontal velocity, $v_x$ and $v_y$, maps. The average supergranule in the simulations with the \f-plane approximation is shown in the middle panel of Fig.~\ref{fig:aveSGdiv}.  As expected, the average supergranule in the non-rotational simulation has purely diverging, radial velocities, and the average supergranule in the  rotational simulation has an azimuthal component with a similar amplitude to the radial component (bottom panel, Fig.~\ref{fig:aveSGdiv}). The average supergranule in the hydrodynamic simulations is qualitatively consistent with the maps of average supergranules on the Sun.  
The radial velocities have similar amplitudes (200-300 ms$^{-1}$), and both the radial and polar components have a similar sine-like profile with a peak about half a supergranule radius from the centre \citep[e.g. see Figs.~12 and 13 in ][and recall that we are simulating the effect from rotation more than 100 times that of the Sun]{Langfellneretal2015_Vorticity}.

\subsection{Flux tube initial condition}\label{sec:fluxtube}

To simulate flux emergence, we placed a magnetic flux tube in the simulation box with an axis aligned in the $\mathbf{\hat{x}}$-direction at $y_c=0$ and $z_c=-11.096$~Mm relative to $z=0$~Mm at the surface. 
The magnetic field is given by
\begin{equation}
    B_x=B_0 \exp[{-\pi B_0 d^2 /\Phi_T}],
\end{equation}
where $d^2=(y-y_c)^2 + (z-z_c)^2$ and $y_c$ and $z_c$ are the $y$-coordinate and $z$-coordinate of the tube axis respectively, the maximum field strength was $B_0=50$~kG, and the total flux was $\Phi_T=10^{21}$~Mx, typical of the observed flux in a small active region at the surface of the Sun. The flux tube had no radial or azimuthal component (twist) and satisfies flux conservation and the solenoidal constraint. The full-width-half-maximum of the radial Gaussian profile was $1.33$~Mm, and we set the magnetic field to zero at a radius of $d=1.71$~Mm, which enclosed 99\% of the total flux.
The width of the flux tube was less than the pressure scale height at this depth.

To maintain pressure balance inside and outside the flux tube, we subtracted the magnetic pressure to make a new initial condition from the non-magnetic snapshot by modifying the pressure to be, 
$p_{\mathrm{new}}=p_0-B^2/2\mu_0$ where $p_0(x,y,z)$ is the pressure before the tube was added.

Because we only want one section of the tube to emerge,  we modify the entropy $s$ in the simulation according to 
\begin{equation}
s_{\mathrm{new}}= s_0 + N \times (s_{\mathrm{cool}}-s_0)
\end{equation}
where $s_0(x,y,z)$ is the entropy before we insert the flux tube,
\begin{equation}
    N = N_1 \frac{B^2}{B_0^2}\left(1-0.2\exp[{-\left(\frac{x}{\sigma}\right)^2}] \right),
\end{equation}
$s_{\mathrm{cool}}=679444350$~K$^{-1}$ (corresponding to very cool material),
$N_1=0.009$, and $\sigma=7.4$~Mm. The parameters $N_1$,  $s_{cool}$ and $\sigma$ were chosen so that only the part of the tube at the horizontal centre of the box ($x=0,y=0$) rises \citep[see ][ for more details on setting the entropy]{Rempel2017}. We note that the horizontal centre of the domain is located in an upflow (see Figs.~\ref{fig:rise0} and \ref{fig:rise110}).
Because we are assuming local thermodynamic equilibrium, specifying $p$ and $s$ specifies the plasma properties. The velocity field $v$ is left unchanged.
Together these specify the flux tube initial condition.

\begin{figure}
    \centering \includegraphics[width=0.99\linewidth]{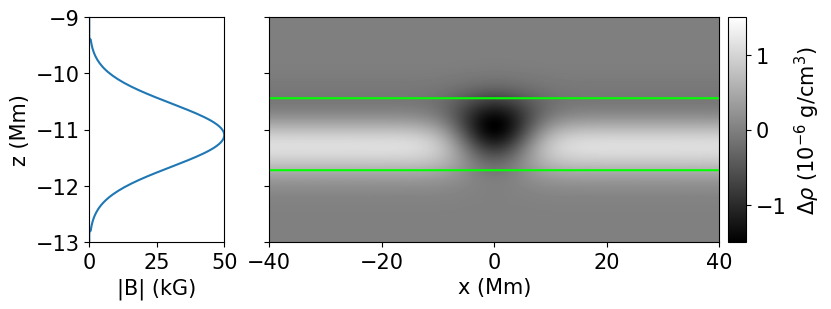}
    \caption{Left: the total magnetic field strength across the centre of the tube as function of height. Right: vertical slice through the centre of the flux tube showing the difference in density between the initial condition with the flux tube and the purely hydrodynamic initial condition. A reduced density (negative) will cause the tube to rise, and an increased density (positive) will cause it to sink. The contour of the magnetic field value of 25~kG is shown in green. The relative density perturbation is constant as a function of height in the background Model~S atmosphere. }
    \label{fig:ictube}
\end{figure}

We begin with this initial condition, and simulate the evolution of the flux tube, first without rotation (Fig.~\ref{fig:rise0}) and then with the \f-plane approximation (Fig.~\ref{fig:rise110}).

\begin{figure}
    \includegraphics[width=0.5\textwidth]{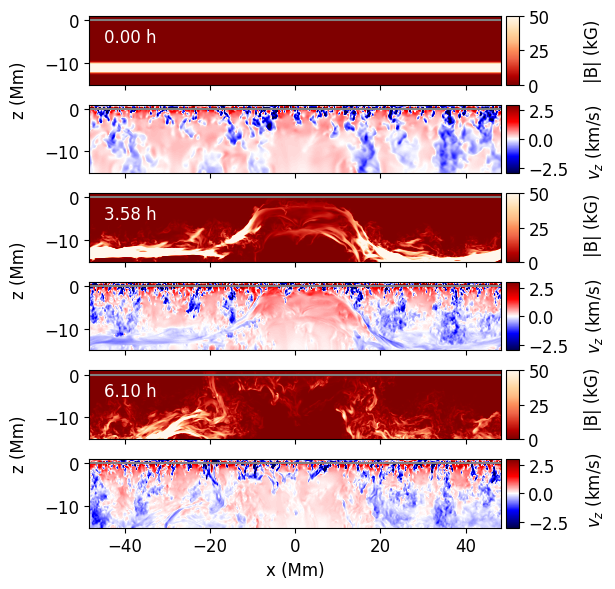}
    \vspace{-0.5cm}
    \caption{Slices through $y=0$ of the magnetic field strength and vertical velocity of the initial condition  (top two panels), and then at two subsequent time steps of the non-rotational ($\Omega = 0 \, \Omega_\odot$) MHD simulation below. The flux tube rises in an arch structure before forming two dominant legs. In the velocity slices, blue indicates downward flow and red indicates upward flow. See electronic movies of the simulation,  \href{https://doi.org/10.17617/3.X57WAX}{flux0video.mp4, here}.}
    \label{fig:rise0}
\end{figure}

\begin{figure}
    \includegraphics[width=0.5\textwidth]{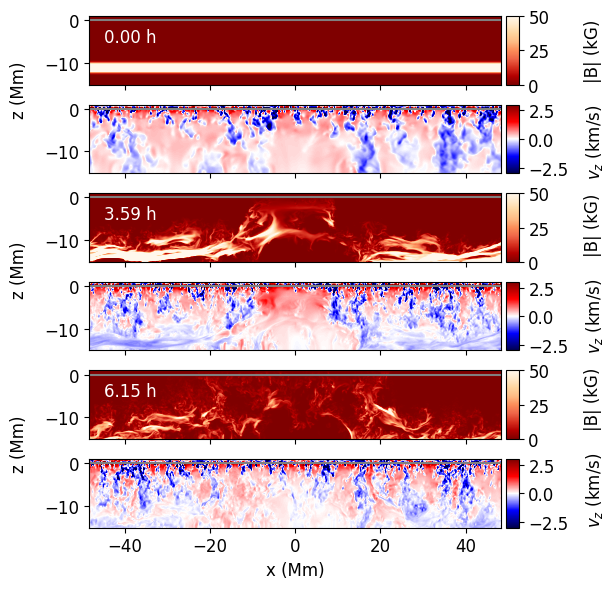}
    \vspace{-0.5cm}
    \caption{Slices through $y=0$ of the magnetic field strength and vertical velocity of the initial condition  (top two panels), and then at two subsequent time steps of the rotational ($\Omega = 110 \, \Omega_\odot$) MHD simulation below. In the velocity slices, blue indicates downward flow and red indicates upward flow. See electronic movies of the simulation,  \href{https://doi.org/10.17617/3.X57WAX}{flux1video.mp4, here}.}
    \label{fig:rise110}
\end{figure}

\begin{figure}
    \centering
    \includegraphics[width=0.99\linewidth]{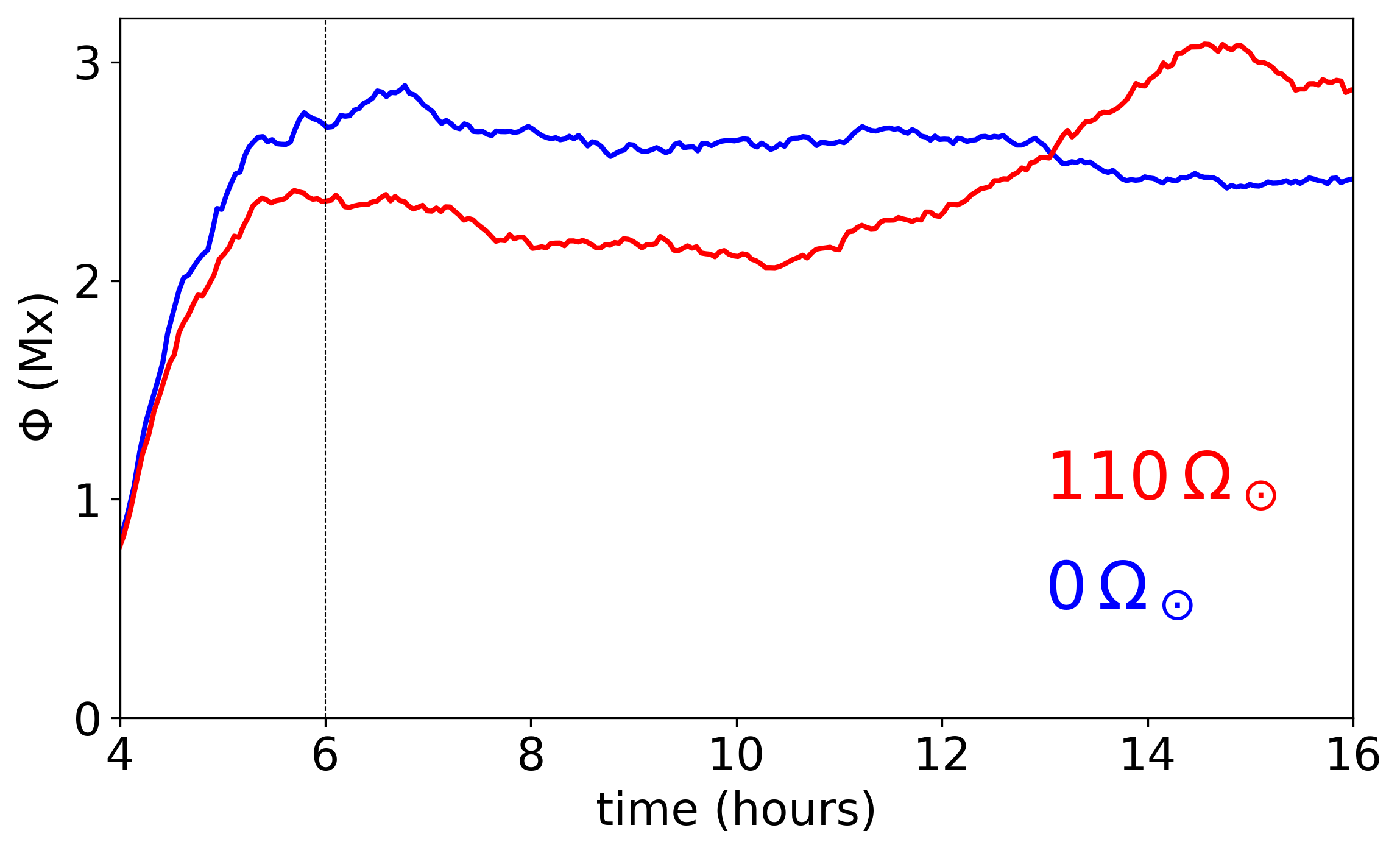}
    \caption{Total unsigned magnetic flux at the surface as a function of time for both non-rotational and rotational MHD simulations. The vertical line indicates the approximate end time of the flux emergence at $6$~hours. }
    \label{fig:absb}
\end{figure}

\begin{figure*}
    \includegraphics[width=0.95\textwidth]{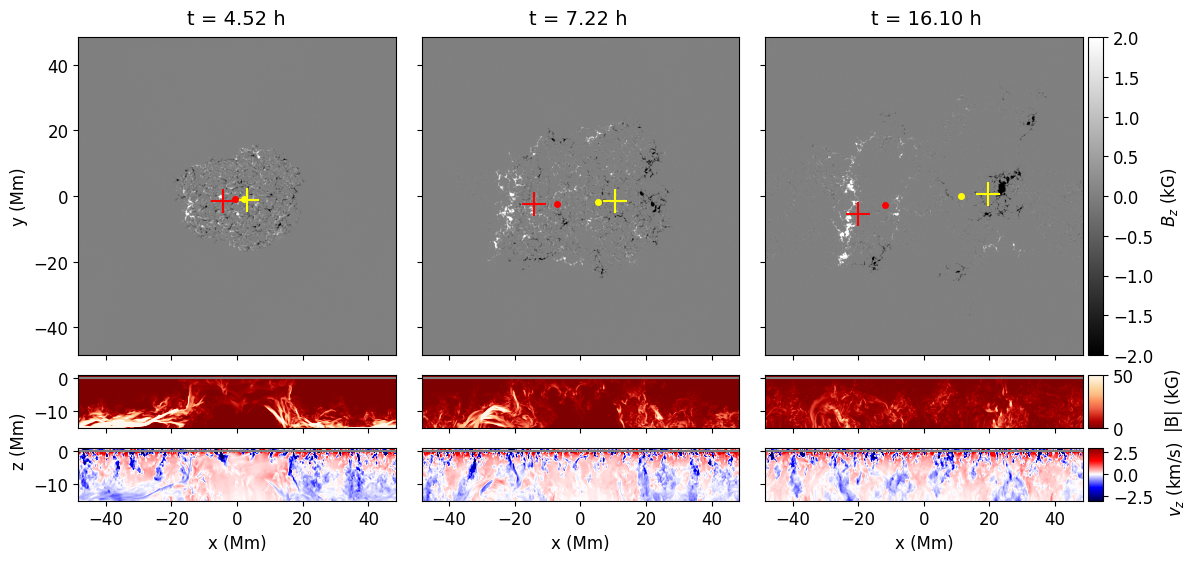}
    \caption{Vertical magnetic field at the surface (greyscale maps) and slices through $y=0$~Mm showing the unsigned magnetic field strength and vertical flows at three different times of the $\Omega = 0 \Omega_\odot$ non-rotational MHD simulation. The red and yellow symbols show the centres of the positive and negative magnetic poles respectively. The crosses indicate the flux-weighted centres of the polarities over a threshold of $|B_z|=800$ G, and the dots indicate the centres without a threshold. See electronic movies of the simulation,  \href{https://doi.org/10.17617/3.X57WAX}{flux0video.mp4, here}.}
    \label{fig:bsurf0}
\end{figure*}

\begin{figure*}  \includegraphics[width=0.95\textwidth]{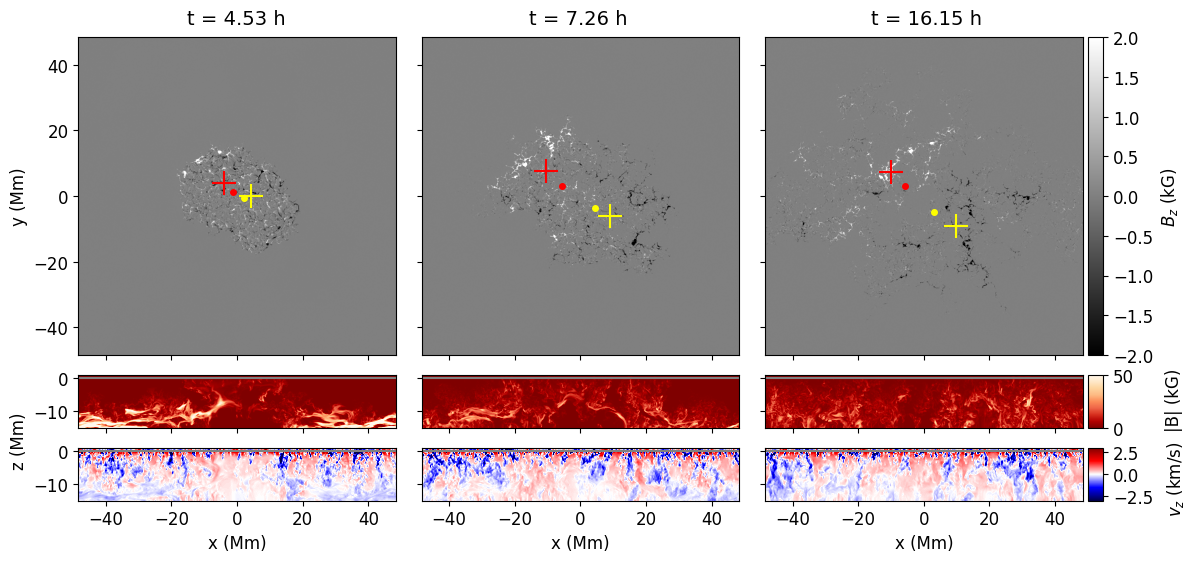}
    \caption{Vertical magnetic field at the surface (greyscale maps) and slices through $y=0$~Mm showing the unsigned magnetic field strength and vertical flows at three different times of the $110 \, \Omega_\odot$ rotational MHD simulation. The red and yellow symbols show the centres of the positive and negative magnetic poles respectively used to calculate the geometric tilt angle, $\gamma_\geom$. The crosses indicate the flux-weighted centres of the polarities over a threshold of $|B_z|=800$ G, and the dots indicate the centres without a threshold. See electronic movies of the simulation,  \href{https://doi.org/10.17617/3.X57WAX}{flux1video.mp4, here}.}
    \label{fig:bsurf110}
\end{figure*}

\subsection{Emergence of the flux tube at the surface}\label{sec:emft}
In both the non-rotational and rotational MHD simulations the emergence process occurs between $t=4$ and $t=6$~hours  (Fig.~\ref{fig:absb}), which is much faster than that observed on the Sun \citep[about 2 days][]{Weberetal2023}.

A general bipole magnetic field structure appeared at the surface at about $t=4$~hours, and coherent polarities with intensity darkening formed at about $t=7$~hours in the non-rotational MHD simulation (Fig.~\ref{fig:bsurf0}) roughly aligned along the flux tube axis (along $y=0$).
The unsigned flux at the surface is about twice the flux in the initial condition (as expected), and comparable to observed active regions at the very beginning of their emergence at the surface \citep[e.g.][]{Schunkeretal2016,Schunkeretal2019}.

\begin{figure}
    \centering
    \includegraphics[width=0.5\textwidth]{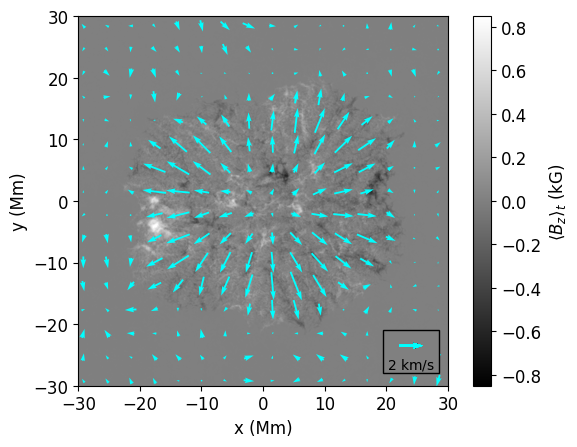}
    \includegraphics[width=0.5\textwidth]{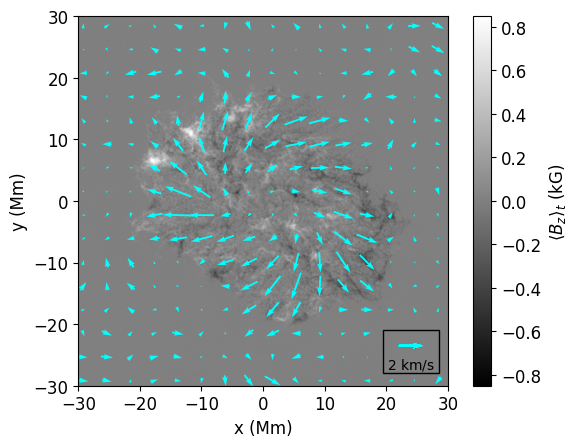}
    \caption{Surface flow velocities (cyan arrows) and vertical magnetic field averaged over time from $t=4.25$~h to $t=5.44$~h for the non-rotational ($0~ \, \Omega_\odot$) hydrodynamic simulation (top) and the rotational ($110~ \, \Omega_\odot$) hydrodynamic simulation (bottom). The velocity has also been spatially smoothed for clarity by a gaussian with standard deviation of $2.4$~Mm.
    Note: this is computed at constant geometrical height, at $z=480~\mathrm{pixels}=15.360$~Mm from the bottom of the box.
    }
    \label{fig:emergenceflows}
\end{figure}

The apex of the rising flux tube is accompanied by a significant horizontal outflow (a positive divergence) at the surface, on the order of 2~km~s$^{-1}$ (Fig.~\ref{fig:emergenceflows}), which we note is inconsistent with observations \citep[e.g.][]{Birchetal2016,Schunker+2024}.
The outflow is due to the fast rise speed of the flux tube (see middle panel of Figs.~\ref{fig:rise0} and \ref{fig:rise110}). A future goal is to develop an initial condition so that the simulation produces a realistic active region that rises slower and does not produce these strong surface outflows.

We note that there are two stages of emergence in the rotational simulation (see Fig.~\ref{fig:absb}). The magnetic flux that reaches the surface in the rotational simulation is initially about $2 \times 10^{21}$~Mx, not substantially less than the non-rotational simulation at about $2.5 \times 10^{21}$~Mx. In the rotational simulation the flux grows again from about 10~hours, peaking at about $2.7 \times 10^{21}$~Mx.
This difference in flux between the simulations may be due to the effects of rotation, or from the non-linear nature of the simulations, however, a number of additional simulations with different convective states would be required to determine that.

\section{Relationship between the motion of the polarities and surface flows}\label{sec:tiltangle}

The positions of the surface polarities begin close together and separate as more flux emerges, as observed for active regions emerging on the Sun.
The fast rise-speed of the flux tube generates outflows (Fig.~\ref{fig:emergenceflows}) that are an order of magnitude greater ($\sim 2$~\kms) than the convective flows ($\sim 100$~\ms) during the emergence phase. Such outflows  are not observed on the Sun \citep{Birchetal2019}.  However, the fast rise speed coupled with the fast rotation rate produces a clear effect of the Coriolis force on the horizontal flows associated with a rising flux tube. In the non-rotational simulation, the polarities emerge at the surface aligned with the flux tube axis (Fig.~\ref{fig:bsurf0}). Towards the end of the simulation a small counter-clockwise tilt develops. In the rotational simulation, the flux tube emerges with a larger tilt angle in the clockwise direction (Fig.~\ref{fig:bsurf110}). 

In this section we measure the geometric tilt angle and compute the component of the tilt angle expected from the Coriolis force acting on the surface horizontal flows.

\subsection{Geometric tilt angle}
To track the location of the polarities, we computed the flux-weighted centre of each polarity,  at the surface as a function of time (see Figs.~\ref{fig:bsurf0} and \ref{fig:bsurf110}). We defined the geometric tilt angle as the angle between the $x$-axis aligned with the flux tube initial condition and the line connecting the two polarities at some time, 
\begin{equation}
\gamma_\geom = - \arctan(\frac{\Delta y_\geom}{\Delta x_\geom}),
\end{equation}
where $\Delta x_\geom$ and $\Delta y_\geom$ are the separation between the polarities (i.e. $\Delta x_\geom = x_- - x_+$ and $\Delta y_\geom = y_- - y_+$) and we defined a positive tilt in the clockwise direction, consistent with the direction of Joy's law in the northern hemisphere on the Sun. 

By design, in the rotational simulations we imposed a rotation rate more than a hundred times faster than the real Sun, and the polarities emerge with a significant tilt in the clockwise direction of about $30^\circ$  (see Fig.~\ref{fig:bsurf110} and Fig.~\ref{fig:emergenceflows}). 
In both the non-rotational and rotational cases most of the flux has emerged by $t=6$~hours (Fig.~\ref{fig:absb}) and, by inspection, any change in tilt angle after that time is not due to significant additional flux emerging from below, but from the motion of existing surface flux. Figure~\ref{fig:separationcomparison} shows that the motion is predominantly in the $y$-direction \citep[consistent with][]{Schunker+2020}, since $\Delta x_\mathrm{g}$  remains roughly constant after emergence. The question is then how much of the motion of the polarities can be explained  by the near-surface horizontal flows.

We note that when measuring the tilt angle in observations it is usually the tilt angle between coherent magnetic structures such as sunspots, and thresholds of magnetic field strength have been used to isolate dominant coherent structures in magnetic field maps \citep[e.g.][]{Schunkeretal2016,Sreedevi+2023}. To demonstrate the equivalence, we include an example of the polarity locations and tilt angle computed only for $|B_z|>800$~G (see Figs.~\ref{fig:bsurf0}, \ref{fig:bsurf110} and \ref{fig:tiltcomparison}) that isolates the small pores that form. The polarity separation is larger in both directions, however the tilt angle is then not significantly different.

\subsection{Tilt angle due to surface horizontal flows}

The geometric positions of the polarities are affected by the motion of the vertical magnetic field due to the horizontal flows, the appearance of vertical field from flux emerging through the surface, and a diffusive component.
In this section we measure the expected contribution to the motion of the polarities from the surface horizontal flows, $\Delta x_\flows$ and $\Delta y_\flows$.  

\begin{figure}
    \centering
    \includegraphics[width=0.99\linewidth]{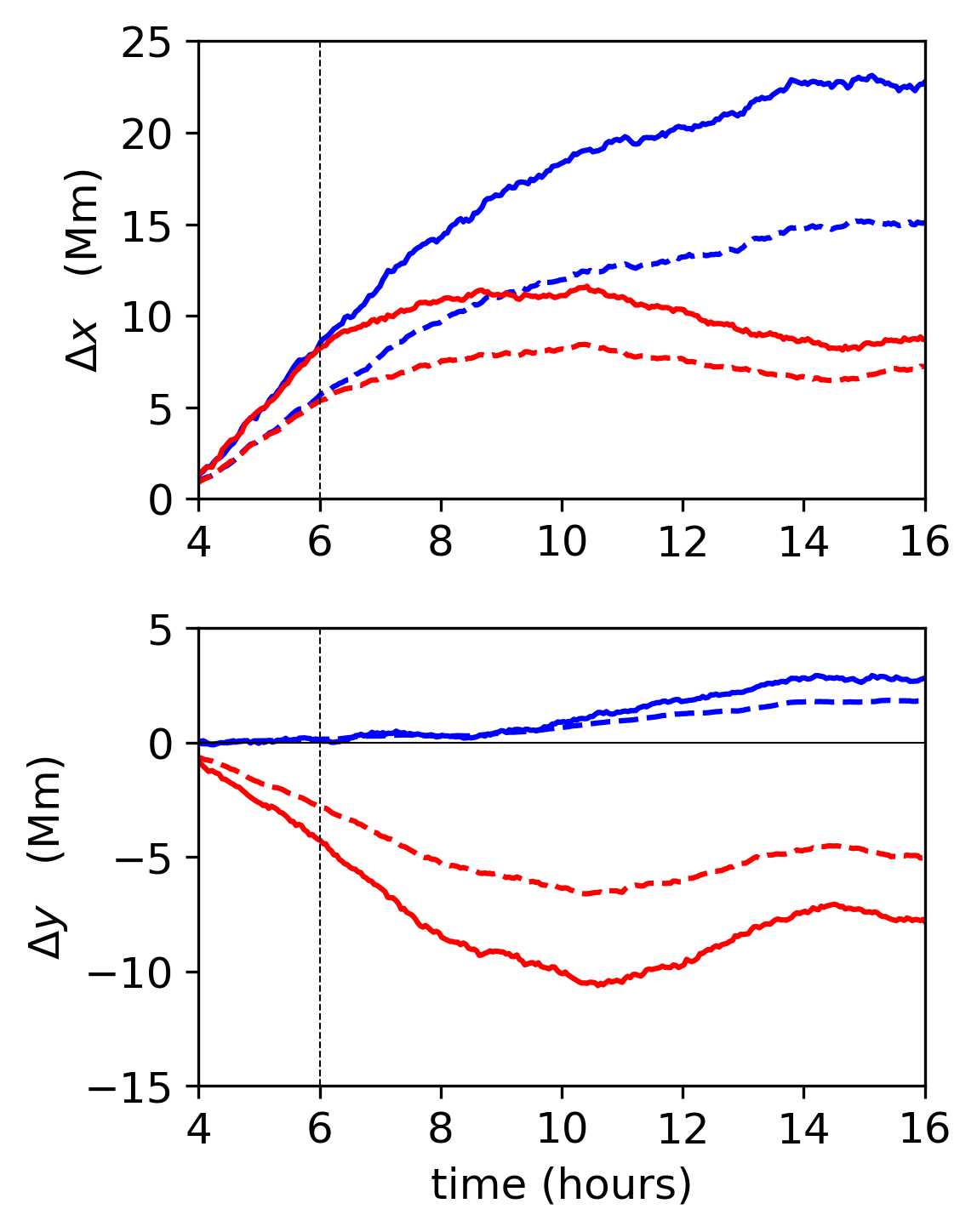}
    \caption{Separation of polarities for the non-rotational simulation (blue) and the rotational simulation (red). 
    Geometric separation without a $B_z$ threshold (solid curves, $\Delta x_\geom $),  and the separation expected from the flows (Eqns.~\ref{eqn:xflowcont} and ~\ref{eqn:yflowcont}, dashed curves, $\Delta x_\flows$). 
    The vertical line at 6~hours is close to the time when most of the magnetic flux has emerged (see Fig.~\ref{fig:absb}). } 
    \label{fig:separationcomparison}
\end{figure}

We compute the separation of the polarities in the $x$-direction due to the flows, $\Delta x_\flows$,  by integrating the horizontal flow velocities weighted by the magnetic field : 
\begin{eqnarray}
 \Delta x_\flows (T) &= \frac{-1}{\langle | B_z(T) | \rangle_A} \int_{t_0}^T \langle v_x B_z \rangle_A \, \mathrm{d}t  \label{eqn:xflowcont} \\
\Delta y_\flows (T) &= \frac{-1}{\langle | B_z(T) | \rangle_A }  \int_{t_0}^T \langle v_y B_z \rangle_A \, \mathrm{d}t,
\label{eqn:yflowcont}
\end{eqnarray}
where the angle brackets indicate the mean over the surface area of the simulation and the integral is cumulative in time  from  $t_0=3$~hours after which there is consistently non-zero magnetic field at the surface (Fig.~\ref{fig:separationcomparison}). 
Our flux tube lies along $y=0$ and the magnetic field is pointed in the positive $x$-direction. Our sign convention is that a positive $\Delta x$ corresponds to the negative polarity lying in the positive $x$ direction relative to the positive polarity, and a positive $\Delta y$ corresponds to the negative polarity lying in the negative $y$ direction relative to the positive polarity, which explains the negative sign in Eqns.~\ref{eqn:xflowcont} and \ref{eqn:yflowcont}. 

\begin{figure}
    \centering
    \includegraphics[width=0.99\linewidth]{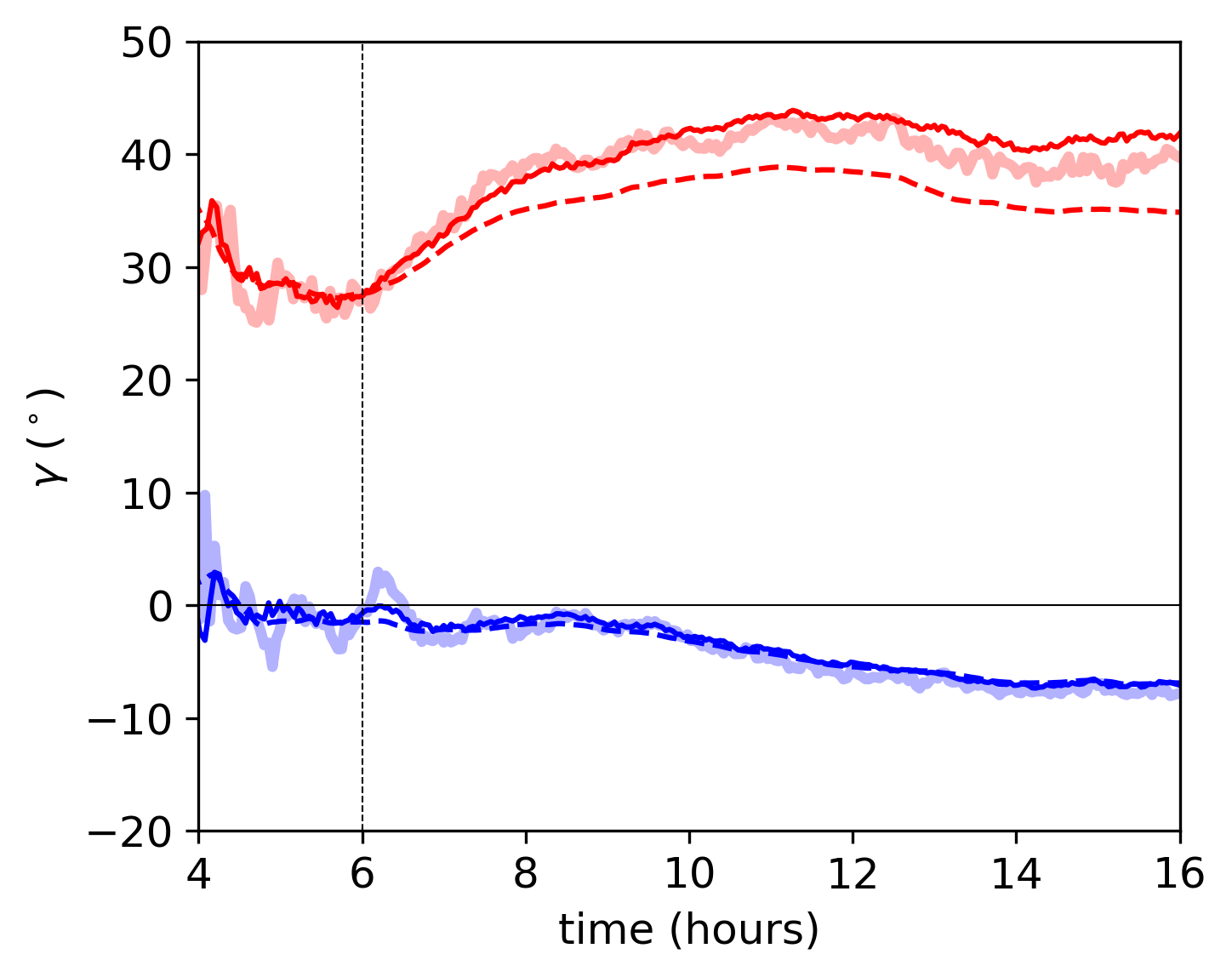}
    \caption{Tilt angle of the polarities for the non-rotational simulation (blue) and the rotational simulation (red). The geometric tilt angle is shown by the solid curves, and the tilt angle expected from the flows is shown by the dashed curves.
    The tilt angle resulting from the location of the polarities with a threshold of $|B_z|>800$~G applied is shown by the thick shaded curves. }
     \label{fig:tiltcomparison}
\end{figure}

The tilt angle imparted by the flows is then
\begin{equation}
\centering
    \gamma_\flows (T) = - \arctan \left( \frac{\Delta y_\flows(T)}{ \Delta x_\flows(T)} \right),
\end{equation}
defined as positive in the clockwise direction. 
 In both the rotational and non-rotational simulations the tilt angle is not significantly different  to the geometric tilt angle (Fig.~\ref{fig:tiltcomparison}), despite the differences in separation components.
Figure~\ref{fig:separationcomparison} shows that the magnitude of the separation due to the horizontal flows is about one third less than the geometric separation, and so there is some remaining component of motion not due to the horizontal flows at the surface.

\subsection{Remaining tilt angle component}\label{sect:rtilt}

We define the difference between the geometric and horizontal flow components  as the remaining component, $\Delta x_\mathrm{r} =\Delta x_\geom - \Delta x_\flows$ and $\Delta y_\mathrm{r} =\Delta y_\geom - \Delta y_\flows$, and the corresponding tilt angle as $\gamma_\mathrm{r} = \gamma_\geom - \gamma_\flows$.
This remaining component of the separation is then due to the emergence of  magnetic field through the surface and diffusion.  On the Sun, the diffusive component is expected to be small, but not necessarily in the simulations.

In the non-rotational simulation, the tilt angle due to the flow component, $\gamma_\flows$, can fully explain the geometric tilt angle (Fig.~\ref{fig:separationcomparison}).
This is true for the rotational simulation only during the emergence phase (from 4 to 6~hours). After the emergence phase the tilt angle from the flows is systematically smaller than the geometric tilt angle by about $\gamma_r \approx 5^\circ$, but within the estimated uncertainty.
The tilt angle in the non-rotational simulation is an estimate of the uncertainty  in the tilt angle measurement, $\delta \gamma \sim 10^\circ$, by buffeting from convective flows \citep[e.g.][]{WangSheeley1989,Will+2024}.

\section{Discussion}\label{sect:conc}

By implementing the \f-plane approximation in the MURaM code, we have shown that the Coriolis force acts on the horizontal flows at the surface, inducing radial vorticity in supergranules.
We simulated a magnetic flux tube rising from 11~Mm beneath the solar surface that formed a magnetic bipole at the surface and generated significant outflows of about 2~km~s$^{-1}$ out to a radius of about 20~Mm from the initial emergence location. 
Our simulations with and without rotation formed small, low-flux magnetic bipoles consistent with the early stages of active region emergence.

By tracking the motion of the two polarities at the surface, we measured the separation and tilt angle as a function of time. 
Without rotation, the polarities emerged aligned with the initial flux tube axis. 
With a rotation rate about one hundred times the Sun's rotation rate, the polarities emerged tilted close to $30^\circ$ in a clockwise direction away from the initial flux tube axis. In both cases the tilt angle remained constant during the emergence process.

The tilt angle of the bipole formed in the rotational simulation, while consistent with Joy's law, is inconsistent with observations in two respects. Firstly, the mean tilt angle of active regions on the Sun is about $\sim 6^\circ$ with an uncertainty a factor of two larger \citep[e.g.][]{Haleetal1919,DasiEspuigetal2010}.
If we apply a scaling argument for the Sun, with a rotation rate   110 times slower,  surface velocities of about 100~\ms (compared to 2000~\ms), and an emergence time of 2~days (compared to 2~hrs), so that the $y$-separation $\Delta y \sim \Omega v_x \Delta t^2$ becomes 
$$
\Delta y_\odot \rightarrow \Delta y \left( \frac{1}{110} \right)  \left( \frac{100 \textrm{\ms}}{2000 \textrm{\ms}} \right) \left( \frac{48~\mathrm{hrs}}{2~\mathrm{hrs}} \right)^2
$$
and then $\gamma_\odot \sim 7^\circ$, which is on the order of Joy's law on the Sun \citep[e.g.][]{Haleetal1919,WangSheeley1989,Will+2024}. 
Secondly, although we have demonstrated that a tilt angle on the order of Joy's Law can be produced in the top 10~Mm of the convection zone, our simulations have not captured the observed time evolution through emergence.
In the observations, active regions on the Sun emerge east-west aligned on average, and the tilt angle develops throughout the emergence process \citep{Schunker+2020}.

Presumably on the Sun, active regions originate from the global toroidal magnetic field oriented in the east-west direction.
In our simulations, the tilt away from the flux tube axis (along the east-west direction) generates a perpendicular component of the magnetic field (north-south direction), and this is consistent with being partly due to the effect of rotation on the near-surface horizontal convective flows. 

We have shown that a bipole can emerge tilted away from the subsurface orientation of the flux tube axis from rotational forces acting on a flux tube as it rises through the near-surface convective flows. The change in tilt angle once the flux has reached the surface is consistent with a component due to the flows and a remaining component, coming from a combination of the vertical velocity acting on the horizontal magnetic field and the effects of numerical diffusion in the simulation.  
Future simulations of flux emergence should improve upon our experiment by simulating a flux tube with a slower rise speed that ideally forms a stable, higher flux active region as well as using a smaller grid size. 
To isolate the effect of the Coriolis force, we chose not to impose a twisted magnetic field in our flux tube, however it may be an important component for the coherency of the flux tube and even for Joy's law \citep[e.g.][]{Schussler1979,LongcopeFisher1996}.
 These simulations are required to interpret the observational flow signatures as active regions emerge and the tilt angle develops.

\begin{acknowledgements}
We acknowledge the sincere and helpful comments from the anonymous referee towards improving this paper.
HS, WRB and LG acknowledge support from the project ``Preparations for PLATO asteroseismology'' DAAD project 57600926.
This research was undertaken with the assistance of resources and services from the National Computational Infrastructure (NCI), which is supported by the Australian Government.
HS is the recipient of an Australian Research Council Future Fellowship Award (project number FT220100330) and this research was funded by this grant and an Australian Government Research Training Program (RTP) Scholarship from the Australian Government.
RHC and LG acknowledge support from the European Research Council (ERC) under the European Union’s Horizon 2020 research and innovation programme (grant agreement No 810218 WHOLESUN).
HS, DIP and WRB  acknowledge the Awabakal people, the traditional custodians of the unceded land on which their research was undertaken at the University of Newcastle.
\end{acknowledgements}

\section*{Data Availability}

The simulations can be fully reproduced following the description in this paper. 
The version of the code including the \f-plane approximation is available in the MURaM repository. The initial conditions for the simulations can be provided on request. The digital data presented in this paper are available through private communication with the authors.



\bibliographystyle{aa}
\bibliography{main} 


\clearpage
\appendix
\section{Validation of the \f-plane approximation in MURaM}\label{app:valfplane}
\begin{figure}
\centering
    \includegraphics[width=0.8\linewidth]{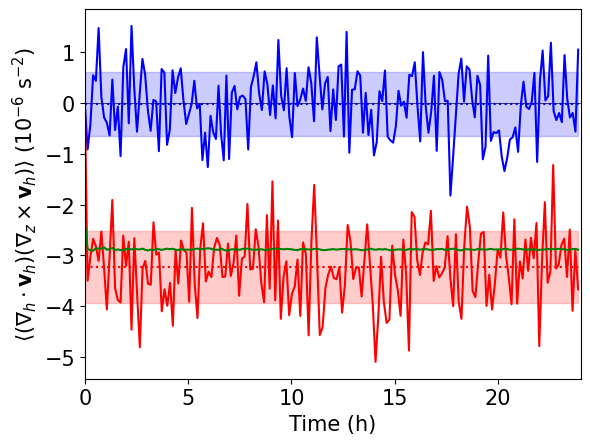}
    \vspace{-0.35cm}
    \caption{Mean product of the vorticity and divergence at the surface for the MURaM hydrodynamic simulations with (red) and without (blue) the \f-plane approximation. The expected analytic value of the product (green, Eqn.~\ref{eqn:an}) is systematically offset but within a standard deviation of the mean of the product for the rotational hydrodynamic simulation with the \f-plane approximation.}
        \label{fig:avep}
\end{figure}

To test the implementation of the \f-plane approximation in the MURaM code, we computed the expected effect on the horizontal flows of a supergranule at the surface. We described the radial profile of velocity  of a supergranule at the surface as a sine function in time and space with a lifetime of $T=\pi/\omega$ and spatial scale of $L=\pi/k$, where $k$ is the spatial scale and  $\omega$ is the temporal scale.  The velocity in the radial direction $v_r$, is given by:
\begin{equation}
    v_r=V_0 \sin(kr) \sin(\omega t)
\end{equation}
for $0\leq t\leq T$, where $r$ is the radius from the centre of the supergranule  and  $V_0$ is the peak velocity. The Coriolis force acts perpendicular to the radial velocity in the plane, so we can write the acceleration due to the Coriolis force as:
\begin{equation}
    \pdv{v_\theta}{t}=-fv_r=-V_0f\sin(kr)\sin(\omega t).
\end{equation}
Integrating with respect to $t$ gives
\begin{equation}
    v_\theta=\frac{V_0f}{\omega}\left(\sin(kr)\cos(\omega t) + C(r)\right).
\end{equation}
If we define $v_\theta=0$ at $t=0$ for all $r$, then
\begin{equation}
    0 = \frac{V_0f}{\omega}\left(\sin(kr) + C(r)\right),
\end{equation}
which leads to
\begin{equation}
    C(r)=-\sin(kr).
\end{equation}

The horizontal flow divergence (in polar coordinates centred on the supergranule) is given by
\begin{equation}
\begin{split}
    \nabla\cdot\mathbf{v}_h&=\frac{1}{r}\pdv{(rv_r)}{r}+\frac{1}{r}\pdv{v_\theta}{\theta} \\
    &=V_0\sin(\omega t)\left[\frac{\sin(kr)}{r}+k\cos(kr)\right],
    \label{eqn:andiv}
\end{split}
\end{equation}
and the vertical component of the vorticity of the horizontal flows is
\begin{equation}
\begin{split}
    (\nabla\times\mathbf{V})_z&=\frac{1}{r}\left(\pdv{(rv_\theta)}{r}-\pdv{v_r}{\theta}\right) \\
    &=\frac{V_0f}{\omega}(1-\cos(\omega t))\left[\frac{\sin(kr)}{r}+k\cos(kr)\right].
    \label{eqn:anvor}
\end{split}
\end{equation}

We weighted the vorticity by the horizontal flow divergence to favour the approximation in the supergranules, by multiplying Eqn.~\ref{eqn:anvor} by Eqn.~\ref{eqn:andiv} to give,
\begin{equation}
\begin{split}
    \mathcal{P} &= (\nabla\cdot\mathbf{v}_h)(\nabla\times\mathbf{v}_h)\\
    &=\frac{V_0^2f}{\omega}(\sin(\omega t) - \sin(\omega t)\cos(\omega t))\left[\frac{\sin(kr)}{r}+k\cos(kr)\right]^2.
\end{split}
\end{equation}
The average product of divergence and vorticity over the lifetime of a supergranule is
\begin{equation}\
\begin{split}
    \langle\mathcal{P}\rangle_T &= \frac{1}{T}\int_0^T \mathcal{P} \text{ d$t$} \\
    &= \frac{2V_0^2 f}{\pi \omega}\left[\frac{\sin(kr)}{r}+k\cos(kr)\right]^2.
    \label{eqn:I}
\end{split}
\end{equation}
Averaging over all surface area of the simulation gives
\begin{equation}
\begin{split}
    \langle\mathcal{P}\rangle &= \frac{1}{A} \int_A \langle\mathcal{P}\rangle_T \text{ d$A$} \\
    &=\frac{1}{\pi \left(\frac{\pi}{2k}\right)^2}\frac{2V_0^2 f}{\pi \omega} \int_{r=0}^{\pi/2k} \int_{\theta=0}^{2\pi} \left[\frac{\sin(kr)}{r}+k\cos(kr)\right]^2 r \text{ d$r$ d$\theta$} \\
    &=\frac{16 V_0^2 k^2 f}{\pi^3 \omega}\int_{r=0}^{\pi/2k} \left[\frac{\sin(kr)}{r}+k\cos(kr)\right]^2 r \text{ d$r$} \\
    &\approx\frac{16 V_0^2 k^2 f}{\pi^3 \omega} (0.12067 \pi^2)
\end{split}
\label{eqn:I2}
\end{equation}
where we solved the integral was numerically. Assuming the lifetime of a supergranule is the time it takes for fluid to cross the diameter once,  $V_0=L/T=\omega/k$, and Eqn.~\ref{eqn:I2} becomes
\begin{equation}
    \langle\mathcal{P}\rangle=\frac{1.93072 V_0 k f}{\pi}.
    \label{eqn:avP}
\end{equation}

To obtain an estimate for the average product independent of the size or lifetime of a supergranule, we expressed the analytic product in terms of the root-mean-square horizontal flow divergence. We derived the mean-square divergence (from Eqn.~\ref{eqn:andiv}) to be
\begin{equation}
\begin{split}
    \langle (\nabla\cdot\mathbf{v}_h)^2 \rangle &= \frac{1}{TA} \int_{t=0}^{\pi/\omega} \int_{\theta=0}^{2\pi} \int_{r=0}^{\pi/2k} (\nabla\cdot\mathbf{v}_h)^2 r \text{ d$r$ d$\theta$ d$t$} \\
    &= 4 V_0^2 k^2 \int_{r=0}^{\pi/2k} \left[\frac{\sin(kr)}{r}+k\cos(kr)\right]^2 r \text{ d$r$} \\
    &= \frac{4 V_0^2 k^2}{\pi^2} (0.12067 \pi^2),
\end{split}
\end{equation}
leading to the root-mean-square divergence given by 
\begin{equation*}
    \sqrt{\langle (\nabla\cdot\mathbf{v}_h)^2 \rangle} = 2V_0 k\sqrt{0.12067}.
\end{equation*}
We measured the root-mean-square divergence of the simulations without the \f-plane approximation, and then substituted that value into  Eqn.~\ref{eqn:avP} to get
\begin{equation}
    \langle\mathcal{P}\rangle=\frac{2.779f}{\pi}\sqrt{\langle (\nabla\cdot\mathbf{v}_h)^2 \rangle}.
    \label{eqn:an}
\end{equation}
Fig.~\ref{fig:avep} shows that this analytic estimate for the spatially averaged product Eqn.~\ref{eqn:an} agrees within the uncertainties with the product computed for the simulations with the \f-plane approximation. Note that the average product is a negative quantity since the value of \f\  is negative due to the direction of rotation.

\end{document}